\documentclass[a4paper,fleqn,usenatbib]{mnras}

\usepackage{newtxtext,newtxmath}
\usepackage[T1]{fontenc}
\usepackage{ae,aecompl}

\usepackage{graphicx}	% Including figure files
\usepackage{amsmath}	% Advanced maths commands
\usepackage{amssymb}	% Extra maths symbols

\usepackage{hyperref}
\hypersetup{
  citecolor=cyan,      % color of links to bibliography
}

\usepackage{graphicx}
\usepackage{color}
\definecolor{orange}{rgb}{1,0.5,0}

\newcommand{\ie}{i.e.,~}
\newcommand{\eg}{e.g.,~}
\newcommand{\cf}{cf.,~}

\title[Constraining high-mass neutron-star mergers]{ The heavier the better: how to constrain mass ratios and
  spins of high-mass neutron-star mergers }

\author[E.R. Most et al.]{Elias R. Most$^{1}$\thanks{emost@itp.uni-frankfurt.de}, Lukas R. Weih$^{1}$,
Luciano Rezzolla$^{1,2}$
\\
$^{1}$ Institut f\"ur Theoretische Physik, Goethe Universit\"at Frankfurt am Main, Germany\\
$^{2}$ School of Mathematics, Trinity College, Dublin 2, Ireland}

\date{Accepted XXX. Received YYY; in original form ZZZ}

\pubyear{2020}

\begin{document}
\label{firstpage}
\maketitle

\begin{abstract}
  The first binary neutron-star merger event, GW170817, and its
  bright electromagnetic counterpart have provided a remarkable amount of
  information. By contrast, the second event, GW190425, with $M_{\rm
    tot}=3.4^{+0.3}_{-0.1}\,M_{\odot}$ and the lack of an electromagnetic
  counterpart, has hardly improved our understanding of neutron-star
  physics. While GW190425 is compatible with a scenario in which the
  merger has lead to a prompt collapse to a black hole and little ejected
  matter to power a counterpart, determining the mass ratio and the
  effective spin $\tilde{\chi}$ of the binary remains
  difficult. This is because gravitational waveforms cannot yet well constrain
  the component spins of the binary. However, since the
  mass of GW190425 is significantly larger than the maximum mass for
  nonrotating neutron stars, $M_{_{\rm TOV}}$, the mass ratio $q$ cannot
  be too small, as the heavier star would not be gravitationally
  stable. Making use of universal relations and a large number of
  equations of state, we provide limits in the $(\tilde{\chi},q)$
  plane for GW190425, namely: $q_{\rm min} \geq 0.38$ and $
  \tilde{\chi}_{\rm max}\leq 0.20$, assuming $M_\mathrm{tot} \simeq 3.4\, M_\odot$.
  Finally, we show how future observations
  of high-mass binaries can provide a lower bound on $M_{_{\rm TOV}}$.
\end{abstract}


\section{Introduction}

After the first detection of gravitational waves from a binary neutron
star system \citep{Abbott2017}, in April of 2019 a second event,
GW190425, was recorded by the LIGO-VIRGO collaboration
\citep{Abbott2020}. For the latter, however, no electromagnetic
counterpart could be detected \citep{Coughlin2020}, which is likely due
to the merger's large distance ($159_{-71}^{+69}\, \rm{Mpc}$), the
corresponding broad sky localisation, or a sign for prompt collapse of
the merger's remnant \citep{Foley2020}. Furthermore, the
total binary mass of $M_{\rm tot}=3.4_{-0.1}^{+0.3}\, M_\odot$ is
unexpectedly high and in contradiction to observed Galactic populations
of neutron stars in binary systems, which have masses of $2.5-2.89\,
M_\odot$ \citep[see, \eg][]{Farrow2019}, thus challenging our knowledge
of its formation channel \citep{Romero-Shaw2020,Safarzadeh2020}. Combined
with only a very weak upper bound on the tidal deformability
$\tilde{\Lambda}< 1000$, these uncertainties have prevented significant
constraints on the equation of state (EOS). Indeed, even the
possibility of classifying this event as a black hole-neutron star
merger is still possible -- albeit less likely
\citep{Kyutoku2020,Han2020}. This is to be contrasted with
the multimessenger event GW170817, which -- thanks to the clear
electromagnetic counterpart and strong limits on the tidal deformability
$\tilde{\Lambda}$ -- has led to a number of constraints on the EOS of
nuclear matter \citep[see][for an incomplete
  list]{Margalit2017,Rezzolla2017,Ruiz2017,Annala2017,Radice2017b,Most2018,Tews2018a,De2018,Abbott2018b,Montana2018,Shibata2019,Koeppel2019}.
However, the unusually high mass of GW190425 does allow us to constrain
the effective spin, $\tilde{\chi}$, and mass ratio, $q$, of the
system. To establish such constraints under the assumption of GW190425
having been a neutron-star binary merger, we make use of universal
relations \citep{Breu2016} for the critical mass and compactness of
rotating neutron stars and combine these relations with a large, publicly
available dataset of $\mathcal{O}(10^6)$ EOSs consistent with the
aforementioned constrains derived from GW170817
\citep{Weih2019_zenodo}. In this way, we can not only constrain the
parameter space of GW190425, but also provide new and robust constraints
for $q$ and $\tilde{\chi}$ that can be used to rule out the most extreme
orbital configurations in future gravitational-wave events of
neutron-star binaries even in the absence of an electromagnetic
counterpart.

\section{The broadbrush picture}
\label{sec:broadbrush}

\begin{figure}
  \centering
  \includegraphics[width=0.4\textwidth]{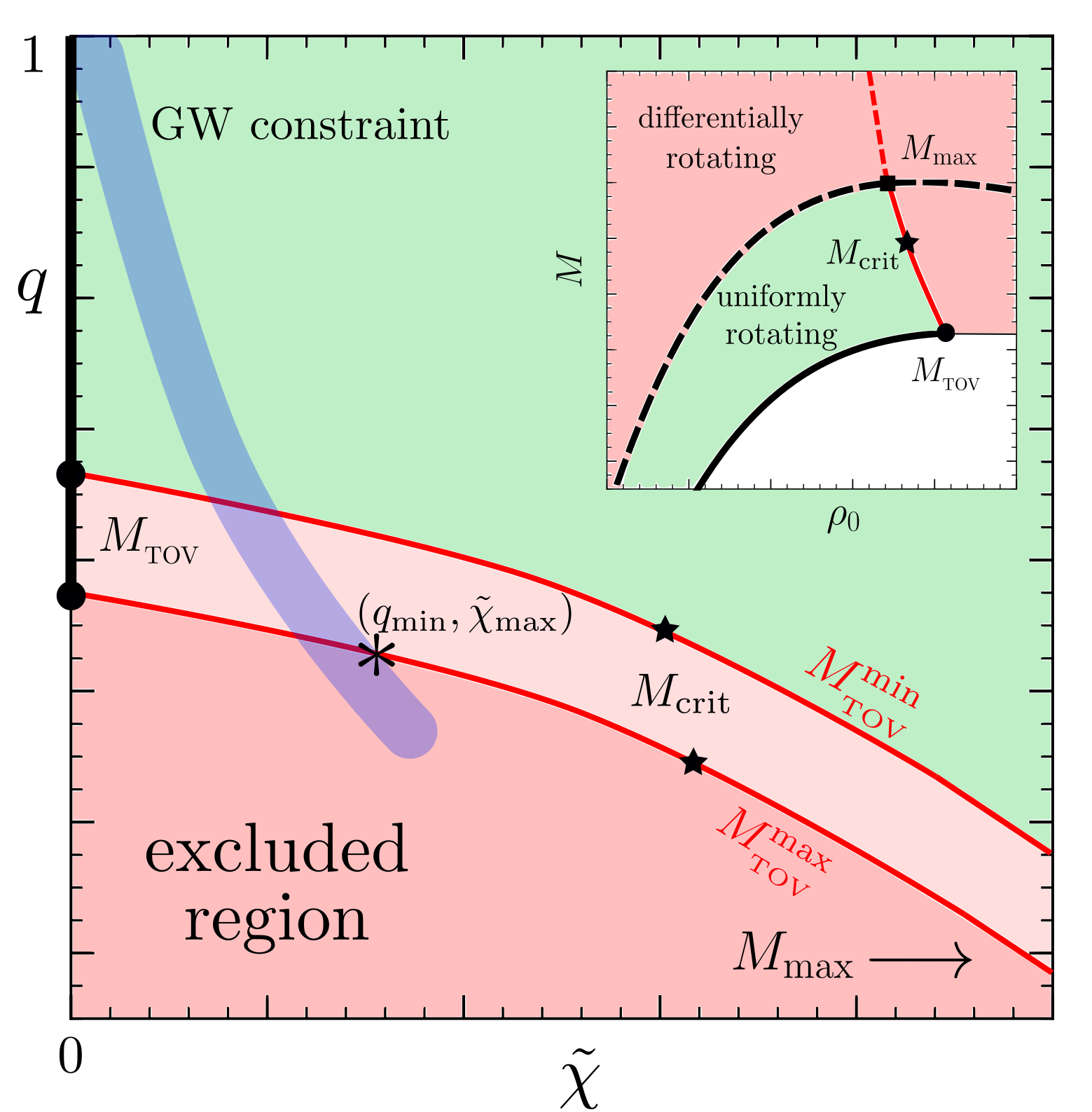}
  \caption{Broadbrush picture of how to constrain the mass ratio $q$ and
    effective dimensionless spin $\tilde{\chi}$ of a high-mass merger.
    The inset reports the same configurations in terms of the mass of the
    star $M$ and its central density $\rho_0$. A detection of a merger
    (blue-shaded area) set constraints on models that are gravitationally
    stable (green-shaded area) and on models that are gravitationally
    unstable (red-shaded area; see Sec. \ref{sec:broadbrush} for
    details).}
  \label{fig:cartoon}
\end{figure}

To illustrate our method for constraining the allowed parameter space for
the mass ratio $q$ and the effective dimensionless spin $\tilde{\chi}$
[see \eqref{eqn:mq12} and \eqref{eqn:chitilde} for definitions] for
high-mass mergers we make use of the schematic diagram shown in
Fig. \ref{fig:cartoon}. We recall that gravitational-wave detections
measure the chirp mass with high precision and thus provide an accurate
estimate for the total mass $M_{\rm tot}$ of the system, whereas the mass
ratio $q$ and dimensionless spin $\tilde{\chi}$ are much less
constrained. Consider therefore a gravitational-wave detection and an
inferred $(\tilde{\chi},q)$ allowed band (shaded-blue area in
Fig. \ref{fig:cartoon}). Such a band will likely include a region of
allowed solutions corresponding to either nonrotating or uniformly
rotating neutron stars (green-shaded area), and an excluded region in
which no such configuration can be found (red-shaded area).  More
precisely, given a neutron-star binary with total mass $M_{\rm tot}$ and
mass ratio $q:=m_2/m_1\leq1$, the component masses of the binary $m_1$
(primary) and $m_2$ (secondary) are given respectively by
\begin{equation}
  \label{eqn:mq12} 
  m_1 = \frac{1}{1+q} M_{\rm tot}\,,
  \qquad
  m_2 = \frac{q}{1+q} M_{\rm tot}\,.
\end{equation}
Hence, assuming some value for the mass ratio $q$ we can then infer the
component masses $m_1$ and $m_2$ of the binary using
Eqs. \eqref{eqn:mq12}, so long as they correspond to gravitationally
stable configurations. If the more massive (primary) star is
nonorotating, then it must have a mass smaller than the maximum mass for
nonrotating configurations, \ie $ m_1 < M_{_{\rm TOV}}$ ($M_{_{\rm TOV}}$
is marked with black circles for different EOSs). On the other hand, if
it is rotationally supported with spin $S_1$, then it must have a mass
smaller than the ``critical mass'' $M_{\rm crit}(S_1)$, that is, the
maximum mass allowed for such a spin, \ie $m_1 < M_{\rm crit}(S_1)$
($M_{\rm crit}$ is marked with black stars). Such a critical mass follows
the neutral-stability line of rotating equilibrium configurations
\citep{Takami:2011, Weih2017} (red solid line in the inset) and is not
straightforward to compute. However, it is well approximated by the
turning-point line \citep{Friedman88}, which is far simpler to compute
and has a maximum value, $M_{\rm max}$, representing the largest mass for
configurations on the mass-shedding limit (dashed black line in the
inset). More importantly, \emph{both} the maximum mass \emph{and} the
critical masses are related to $M_{_{\rm TOV}}$ through universal
relations, \ie $M_\mathrm{crit} \leq M_{\rm max} \simeq 1.2\,
M_{_{\mathrm{TOV}}}$ \citep{Breu2016}; $M_{\rm max}$ is marked with black
a filled square in the inset. In essence, therefore, for the primary to
be gravitationally stable, its position in the $(\tilde{\chi},q)$ plane
will have to be in the green-shaded region; outside this region, the
primary is either unstable or differentially rotating.

As a result, for any value of the dimensionless spin $\tilde{\chi}$, the
green-shaded area is bounded from below by a ``critical mass ratio''
\begin{align}
  q_{\rm{crit}} \left( \tilde{\chi}, M_{\rm tot} \right)
  := \frac{M_{\rm{tot}}}{M_{\rm{crit}}(\tilde{\chi})} - 1\,,
  \label{eqn:q_crit}
\end{align}
set only by the critical mass at that dimensionless spin an by the total
mass of the binary. Note that after some (reasonable) assumption about
the spin distribution in the binary, expression \eqref{eqn:q_crit} can
effectively be cast only in terms of the dimensionless spin of the
primary $\chi_1:=S_1/m^2_1$. This follows directly from the definition of
the dimensionless spin
\begin{equation}
  \label{eqn:chitilde}
  \tilde{\chi} :=\frac{m_1 \chi_1 + m_2 \chi_2 }{m_2 + m_1} =
  \frac{\chi_1}{1+q}\left( 1 + q\frac{\chi_2}{\chi_1} \right) =
  \frac{\chi_1}{1+q}\left( 1 + \frac{1}{q}\frac{S_2}{S_1} \right)\,,
  \end{equation}
so that an estimate of the ratio of the two components' spin,
$\chi_1/\chi_2$ ($S_2$ and $\chi_2$ are the spin and dimensionless spin
of the secondary, respectively) allows one to express $q_{\rm{crit}} =
q_{\rm{crit}} (\chi_1, M_{\rm tot})$.

Due to the uncertainty in the EOS of nuclear matter, this exclusion
boundary is not known exactly and is therefore represented by the
light-red shaded area in Fig. \ref{fig:cartoon}. However, by computing
this boundary for a large dataset of $\mathcal{O}(10^6)$ EOSs, we can
find a lower limit below which an excluded region can be defined
safely. This lower limit is given by the EOSs with the highest
$M_{_{\rm{TOV}}}$, \ie $M_{_{\rm{TOV}}}^{\rm{max}}$ for a given
dimensionless spin, thus defining the minimum critical mass ratio
$q_{\min}:= q_{\rm{crit}} \left( \tilde{\chi}, M_{\rm tot}, M_{_{\rm
    TOV}}^{\rm{max}} \right)$. In turn, $q_{\min}$ provides an upper
limit on $\tilde{\chi}$, \ie $\tilde{\chi}_{\rm max}$ (asterisk in
Fig. \ref{fig:cartoon}).

In what follows, we apply the general considerations made so far to the
specific case of GW190425 and discuss how we can set new limits on
$M_{_{\mathrm{TOV}}}$ with future detections of high-mass binaries.

\section{Universal relations and fits}

While the properties of neutron stars depend on their composition and
hence on the still unknown EOS, some of their bulk properties, such as
the moment of inertia or the quadrupole moment, have been shown to obey
quasi-universal relations (see \citealt{Yagi2017} for a review, but also
\citealt{Haskell2014} and \citealt{Doneva2014a} for examples of when
these relations break). The appeal of quasi-universality is that they
allow to make predictions and place constraints independent of the
uncertainty in the EOS. For this reason, we will make use of them in
order to compute a solid estimate of the exclusion line
\eqref{eqn:q_crit}.

According to \cite{Breu2016}, the critical mass $M_{\rm crit}$ (\ie the
mass along the red solid line in the inset of Fig. \ref{fig:cartoon})
fulfils a quasi-universal relation
\begin{align}
  M_{\rm crit}(\chi_{\rm{crit}}, \chi_\mathrm{kep}, M_{_{\rm TOV}}) := M_{_{\rm TOV}} \left( 1 + 
	a_1 \left( \frac{\chi_{\rm{crit}}}{\chi_{\rm Kep}} \right)^2
	+a_2 \left( \frac{\chi_{\rm{crit}}}{\chi_{\rm Kep}} \right)^4
  \right)\,,
  \label{eqn:M_crit}
\end{align}
where $a_1 = 0.132$, $a_2 = 0.071$ and $\chi_{\rm Kep}$ is the largest
possible dimensionless spin for the primary star at the mass-shedding
limit. At this limit, $M_{\rm crit}(\chi_\mathrm{crit} =
\chi_{\rm{Kep}})=:M_{\rm max}=(1.203\pm0.022)\,M_{_{\rm TOV}}$ and
$\chi_{\rm{Kep}}$ is related to the compactness $\mathcal{C}_{_{\rm TOV}}
:= M_{_{\rm TOV}}/ R_{_{\rm TOV}}$, where $R_{_{\rm TOV}}$ is the radius
of the nonrotating neutron star configuration with the maximum mass
$M_{_{\rm TOV}}$. In particular, starting from the analysis of
\citet{Breu2016} and \citet{Koliogiannis2020}, \citet{Shao2020} found
that for large compactnesses the dimensionless Keplerian angular momentum
shows a weak but linear dependence on $\sqrt{\mathcal{C}}$. We therefore
model this behaviour as
\begin{align}
  \chi_{\rm Kep} \simeq \frac{\alpha_1}{\sqrt{\mathcal{C}_{_{\rm TOV}}}} + \alpha_2
  \sqrt{\mathcal{C}_{_{\rm TOV}}}\,,
  \label{eqn:chik}
\end{align}
where the first term was initially suggested by \cite{Breu2016} and the
second one by \citet{Shao2020}. A direct fit to the data of
\cite{Breu2016} and \citet{Rezzolla2017} yields $\alpha_1= 0.045 \pm
0.021$ and $\alpha_2= 1.112 \pm 0.072$. We should note that since
$\alpha_1 \ll \alpha_2$, \citealt{Shao2020} have taken $\alpha_1=0$ as a
first approximation, although $\alpha_1\neq 0$ is important to recover
the Newtonian limit. Furthermore, because the dependence is weak and the
scatter in the data large, setting $\chi_{\rm Kep} \simeq {\rm
  const.}=0.682$ provides a remarkably good approximation. In addition,
it hints the intriguing suggestion that -- in analogy to black holes -- a
quasi-universal upper limit exists for the dimensionless spin of
uniformly rotating compact stars.

%Using this relation, we can express the
%critical mass of a rotating star as
%
%\begin{align}
%  M_{\rm crit} = M_{_{\rm TOV}} \left( 1 
%  +\frac{a_1}{a_3^2} \mathcal{C}_{_{\rm TOV}} \chi^2 
%  +\frac{a_2}{a_3^4} \mathcal{C}_{_{\rm TOV}}^2 \chi^4 
%  \right)\,,
%  \label{eqn:Mcrit_ctov}
%\end{align}
%
By inverting Eq. \eqref{eqn:M_crit}, we can write the critical
dimensionless spin as
\begin{align}
  \chi_{\rm crit} = \chi_{\rm kep} \left( \mathcal{C}_{_{\rm TOV}} \right)
  \sqrt{-\frac{a_1}{2 a_2} + \sqrt{\frac{a_1^2}{4 a_2^2}
  +\frac{1}{a_2} \left( \frac{M_{\rm crit}}{M_{_{\rm TOV}}} -1 \right)}}\,.
  \label{eqn:chi_crit}
\end{align}
Stated differently, Eq. \eqref{eqn:chi_crit} provides an analytic
expression for the dimensionless spin of stellar models along the stability
line of uniformly rotating configurations (red solid line in
Fig. \ref{fig:cartoon}). This quantity is only a function of $M_{_{\rm
    TOV}}$ and $R_{_{\rm TOV}}$, which are, however, unknown since a
precise EOS of neutron stars is still undetermined. Furthermore, since
$\chi_{\rm crit}\propto \mathcal{C}_{_{\rm TOV}}^{1/2}$, to derive a
lower limit for the dimensionless spin $\chi_{\rm crit}$ needed to
support a given critical mass $M_{\rm crit}$, we obviously need to
provide a lower bound for $\mathcal{C}_{_{\rm TOV}}$, namely,
$\mathcal{C}^{\rm min}_{_{\rm TOV}}$.

Luckily, the multimessenger detection of GW170817 has provided several
constraints on the EOS and the corresponding radii
\citep[\eg][]{Annala2017,Most2018,Abbott2018b,Weih2019} and maximum mass
\citep{Margalit2017,Rezzolla2017,Ruiz2017,Shibata2019} of neutron stars.
In particular, exploiting this detection, \cite{Most2018} and
\cite{Weih2019} have built a large dataset of $\mathcal{O}\left( 10^6
\right)$ EOSs satisfying recent constraints from chiral effective-field
theory \citep{Drischler2016}, tidal deformabilities
\citep{Radice2017b,Abbott2018b, Kiuchi2019}, \ie $280 < \Lambda_{1.4} <
560$ for a $1.4\, M_\odot$ star, and the maximum mass of nonrotating
neutron stars \citep{Margalit2017, Rezzolla2017,Ruiz2017, Shibata2019},
\ie $1.97 < M_{_{\rm TOV}} / M_\odot \lesssim 2.3$. We remark that these
upper bounds on the maximum mass assume the formation of a black hole,
which is plausible given the follow-up observations consistent with a
structured jet \citep[see, \eg][]{Hajela2019}, or the constraints on the
overproduction of blue ejected matter \citep[see, \eg][]{Gill2019}.
However, if a black hole was not formed in GW179817, the maximum mass could
be $M_{_{\rm TOV}} \gtrsim 2.4\, M_\odot$ \citep{Ai2019}. 

Using the before mentioned
dataset, we can find (and fit) an lower bound for $\mathcal{C}_{_{\rm
    TOV}}^{\rm min}$ in terms of the maximum mass for nonrotating stars
$M_{_{\rm TOV}}$ as the latter varies across the various EOSs. In this
way, we derive
\begin{align}
  \sqrt{\mathcal{C}_{_{\rm TOV}}^{\rm min}} =
  c_1 M_{_{\rm TOV}} +
  c_2 M^2_{_{\rm TOV}} +
  c_3 M^3_{_{\rm TOV}}\,,
\label{eqn:C_TOV_max}
\end{align}
where $c_1=0.482$, $c_2=-0.174$, $c_3 = 0.027$. Note that
\eqref{eqn:C_TOV_max} can also be seen as a fit to the upper bound for
$R_{_{\rm TOV}}$ and that, using as upper bound for the maximum mass the
value $M^{\max}_{_\mathrm{TOV}}=2.3\, M_\odot$, we obtain
$\mathcal{C}_{_{\rm TOV}}^{\rm min} \left( M^{\max}_{_\mathrm{TOV}}
\right) = 0.265$, corresponding to a radius
$R_{_{\mathrm{TOV}}}^{\mathrm{max}} = 12.81\,\mathrm{km}$.

At this point, using all the EOSs in our datasets, we can define the
maximum mass that a rotating star with critical dimensionless spin can
support when $\chi_\mathrm{kep} = \chi_\mathrm{kep} \left(
\mathcal{C}_{_{\rm TOV}}^{\rm min} \right) =:
\chi_\mathrm{kep}^\mathrm{min}$ and its mass is $M_{\rm crit}^{\rm max} =
M_{\rm crit} \left( \chi_\mathrm{crit}, \chi_\mathrm{kep}^\mathrm{min}
\left( M_{_\mathrm{TOV}} \right), M_{_\mathrm{TOV}} \right)$. To this
scope, all we need to do is to insert Eq. (\ref{eqn:chik}) in
Eq. (\ref{eqn:M_crit}), with $\mathcal{C}_{_{\mathrm{TOV}}} =
\mathcal{C}^{\rm min}_{_{\mathrm{TOV}}}$ as given by
Eq. \eqref{eqn:C_TOV_max}\footnote{Note that we use $\mathcal{C}^{\rm
    min}_{_{\mathrm{TOV}}}$ and not $\mathcal{C}_{_{\mathrm{TOV}}}$ since
  we are searching for an upper limit and $M_{\rm crit}(\mathcal{C}^{\rm
    min}_{_{\mathrm{TOV}}}, M_{_{\mathrm{TOV}}}, \chi_{\rm crit}) \geq
  M_{\rm crit}(\mathcal{C}_{_{\mathrm{TOV}}}, M_{_{\mathrm{TOV}}},
  \chi_{\rm crit})$ in the dataset.}.

\section{Application to GW190425 and future events}
\label{sec:results}

\begin{figure*}
  \centering
  \includegraphics[width=0.9\textwidth]{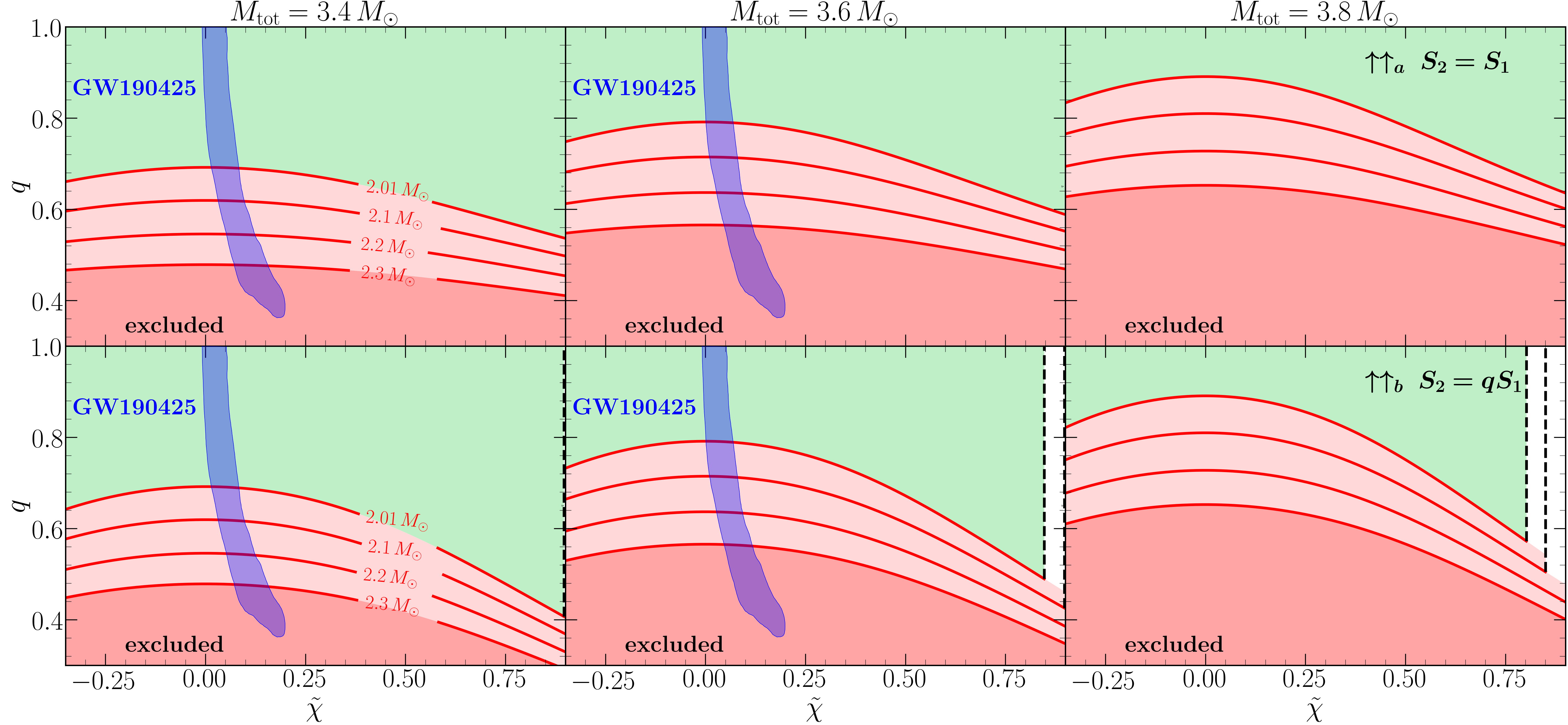}
  \caption{Exclusion lines on the binary mass ratio $q$ and mass weighted
    spin $\tilde{\chi}$ for high-mass binary neutron star mergers with
    total masses $M_\mathrm{tot}/M_\odot = \left( 3.4,3.6,3.8 \right)$
    and aligned spins. Similar to Fig. \ref{fig:cartoon}, the green areas
    represent the allowed parameter space of stable models, while the
    dark-red (excluded) areas do not contain stable binary
    configurations. The exclusion line is determined by the maximum mass
    $M_{_{\mathrm{TOV}}}$, whose uncertainty translates to the light-red
    shaded areas. The observationally inferred parameter range for
    GW190425 is indicated by the blue-shaded area. }
  \label{fig:GW190425_const}
\end{figure*}

\begin{figure*}
  \centering
  \includegraphics[width=0.9\textwidth]{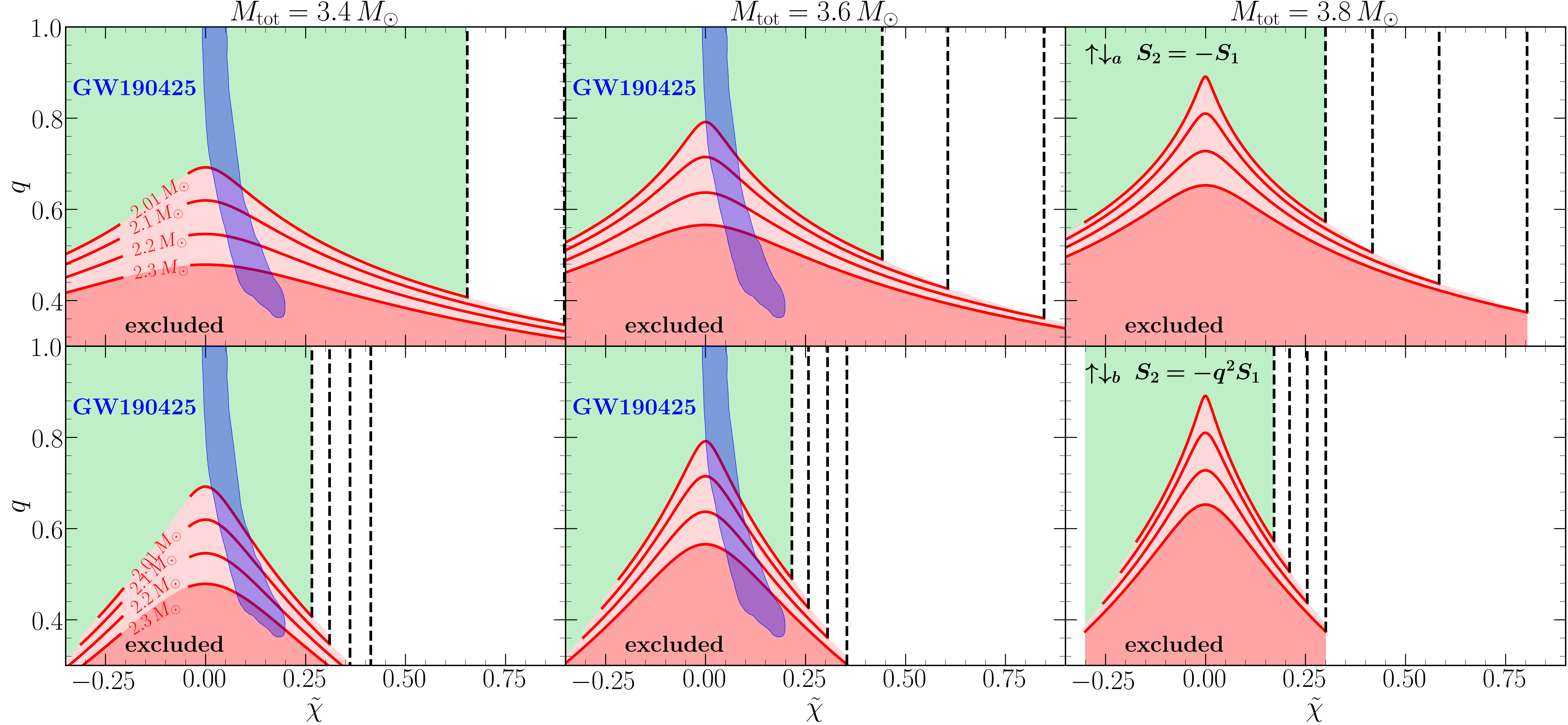}
  \caption{Same as Fig. \ref{fig:GW190425_const} but for antialigned spin
    configurations. The vertical black dashed line refers to the models
    on the mass-shedding limit (\cf Fig. \ref{fig:cartoon}); the
    properties of the models beyond these lines depend on the EOS and could
    either be stable or unstable.}
  \label{fig:GW190425_const_anti}
\end{figure*}

We now apply the methodology illustrated so far to the specific case of
GW190425 and in this way derive constraints on its mass ratio $q$ and
dimensionless spin $\tilde{\chi}$. As outlined in
Sec. \ref{sec:broadbrush}, we need to compute the minimal mass ratio
$q_\mathrm{crit}$ using Eq. \eqref{eqn:q_crit}, together with the
expression for the critical mass ${M}_{\rm crit}$ given by
Eq. \eqref{eqn:M_crit}. Additionally, we need to provide an \emph{ansatz}
for the relation between the two spins $S_1, S_2$, in order to express
$\tilde{\chi} = \tilde{\chi} \left(\chi_1,q\right)$. We therefore consider
four different plausible aligned ($\uparrow\uparrow$) or antialigned
($\uparrow\downarrow$) spin configurations, although it is
straightforward to consider other cases as a function of $q$:
\begin{itemize}
\item[$\uparrow\uparrow_a:$] \quad \makebox[2cm]{$S_2 = S_1 $\hfill}
  $\longleftrightarrow \hspace{1cm} \tilde{\chi}=\chi_1 / q $
\item[$\uparrow\uparrow_b:$] \quad \makebox[2cm]{$S_2 = qS_1 $\hfill}
  $\longleftrightarrow \hspace{1cm} \tilde{\chi}=2\chi_1/(1+q)$
\item[$\uparrow\downarrow_a:$] \quad \makebox[2cm]{$S_2 = -S_1$\hfill}
  $\longleftrightarrow \hspace{1cm} \tilde{\chi}=\chi_1(1-1/q)/(1+q) $
\item[$\uparrow\downarrow_b:$] \quad \makebox[2cm]{$S_2 = -q^2 S_1$\hfill}
  $\longleftrightarrow \hspace{1cm} \tilde{\chi}=\chi_1(1-q)/(1+q) $
\end{itemize}
Note that $\tilde{\chi}=0$ for $S_2 = -q S_1$ or, equivalently, $\chi_2 =
- \chi_1/q$, and that our choices of $\tilde{\chi}$ are deliberately kept
simple to illustrate the power of the method. In particular, all of them
are constructed such that $\tilde{\chi} = 0$ when $\chi_1 =0$; more
elaborate choices can trivially be incorporated into our approach, but
are beyond the scope of this Letter.

Using Eq. \eqref{eqn:M_crit} for the definition of $M_{\rm crit}^{\max}$,
we can then compute the allowed range in mass ratios $q >
q_{\mathrm{crit}}$ as given by Eq. \eqref{eqn:q_crit}. The resulting
parameter space in the $(\tilde{\chi},q)$ plane is shown in
Figs. \ref{fig:GW190425_const} and \ref{fig:GW190425_const_anti} using
the same colour-coding and line conventions introduced in
Fig. \ref{fig:cartoon}.  Overall, we show results for three different
values of the total mass, \ie $M_\mathrm{tot} = 3.4,\,3.6$ and $3.8\,
M_\odot$ (left, middle and right panels respectively) , where $3.4\,
M_\odot$ applies most closely to GW190425. Clearly, the red solid lines
represent the contours separating the allowed/excluded regions for
different choices of $M_{_{\rm TOV}} \in \left[ 2.01;2.3
  \right]\,M_{\odot}$.

Concentrating first on the case of GW190425 with high-spin prior
$\tilde{\chi} < 0.95$ \citep{Abbott2020} (blue-shaded region, left panels
in Figs. \ref{fig:GW190425_const} and \ref{fig:GW190425_const_anti}), we
can appreciate that the constraints from GW190425 do fall in excluded
regions. Stated differently, below a certain mass ratio, no binary
configurations are possible in which the stars can be supported through
rotation and are compatible with the observations. This threshold does
depend on the choice of spin configuration, although in the specific case
of GW190425 only weakly, since the inferred dimensionless effective spins
are small, \ie $\tilde{\chi} < 0.2$. Furthermore, when considering
different values of the maximum mass $M_{_{\rm TOV}}$, this threshold is
present even when taking an upper value of $M_{_{\rm TOV}}\simeq
2.3\,M_{\odot}$, as suggested by \citet{Rezzolla2017} and
\citet{Shibata2019}.

Overall, in the case of GW190425 we can conclude that mass ratios $q
\lesssim 0.5$ are robustly excluded across the different choices of spin
configurations. This exclusion is almost independent of the value of
$\tilde{\chi} <0.2$, since the mass weighted spin is too low for rotation
to significantly increase the supported maximum mass and thus
$M_\mathrm{crit} \simeq M_{_{\rm TOV}}$. To be more precise, we find that
$q_{\rm min} \gtrsim 0.46\ (0.38)$ for aligned (antialigned)
configurations.  Similarly, we can set upper limits on the dimensionless
spin, with $\tilde{\chi}_{\rm max} \lesssim 0.16\ (0.20)$ for aligned
(antialigned) configurations; the corresponding constraints when
considering a mass of $3.6\,M_{\odot}$ for GW190425 are: $q_{\rm min}
\gtrsim 0.56\ (0.46)$ and $\tilde{\chi}_{\rm max} \lesssim
0.12\ (0.17)$. We note that \citet{Foley2020} have recently presented
similar constraints for GW190425 applying a different line of
arguments. In particular, assuming a Gaussian distribution of
$M_{_{\mathrm{TOV}}} = 2.1 \pm 0.12 \, M_\odot$ at $2\sigma$ and prior of
$\chi_1 < 0.4$ chosen to include stellar models computed with a selection
of EOSs, they provide a lower limit $q > 0.53$ at $3\sigma$ for GW190425,
which is in good agreement with our findings.

We next discuss how these results change when considering hypothetical
higher-mass binaries (\ie $M_{\rm tot}=3.6, 3.8\,M_{\odot}$ in the mid
and right panels in Figs. \ref{fig:GW190425_const} and
\ref{fig:GW190425_const_anti}), noting that quite generically the
red-shaded bands of rotationally unstable models, and thus the exclusion
regions, shift upwards [\cf Eq. \eqref{eqn:q_crit}]. For aligned
configurations, the rotational corrections are very small (the red solid
lines are almost horizontal in the middle and right panels of
Fig. \ref{fig:GW190425_const_anti}), so that for $M_{_{\rm TOV}}
=2.3\,M_{\odot}$ we can set nonrotating (\ie $\tilde{\chi}=0$) cutoffs of
$q \gtrsim 0.4$ and $q \gtrsim 0.5$ for the aligned cases
$\uparrow\uparrow_a$ and $\uparrow\uparrow_b$, respectively. In the case
of misaligned spins, on the other hand, the spin effects are much more
pronounced and the exclusion regions move up significantly. As a result,
we can set nonrotating cutoffs of $q \gtrsim 0.56$ and $q \gtrsim 0.65$
for the antialigned cases $\uparrow\downarrow_a$ and
$\uparrow\downarrow_b$, respectively. Interestingly, even for small
dimensionless spins, the degeneracy in $M_{_{\mathrm{TOV}}}$ is removed
with increasing spin and slightly more asymmetric binaries with $q<0.4$
begin to be allowed, albeit at very high individual spins.

Most prominently, shown as white-shaded ares in
Fig. \ref{fig:GW190425_const_anti} are the early terminations of the
regions of rotationally supported neutron stars. This follows directly
from the upper limit of $\chi_\mathrm{Kep} \geq \chi_\mathrm{crit}$ on
the dimensionless spin of rotating stars with maximum masses
$M_\mathrm{crit}$. Since the antialigned cases have a partial
cancellation of the individual component spins, $\chi_1$ will reach
$\chi_\mathrm{Kep}$ already for $\tilde{\chi} < 0.6$. As a trivial
consequence this also implies that there is an exclusion region for
$\chi_1 > \chi_\mathrm{Kep}$ and hence also an upper limit on
$\tilde{\chi} < \tilde{\chi}_{\mathrm{kep}}$, as marked in Fig.
\ref{fig:GW190425_const_anti} with vertical black dashed lines. Note that
the stability properties of the models beyond these lines depend on the
EOS and could lead to either green- or red-shaded regions and to complex
boundaries. For simplicity, and because it is irrelevant for our
discussion here, we simply use a white shading.

\begin{figure}
  \centering
  \includegraphics[width=0.45\textwidth]{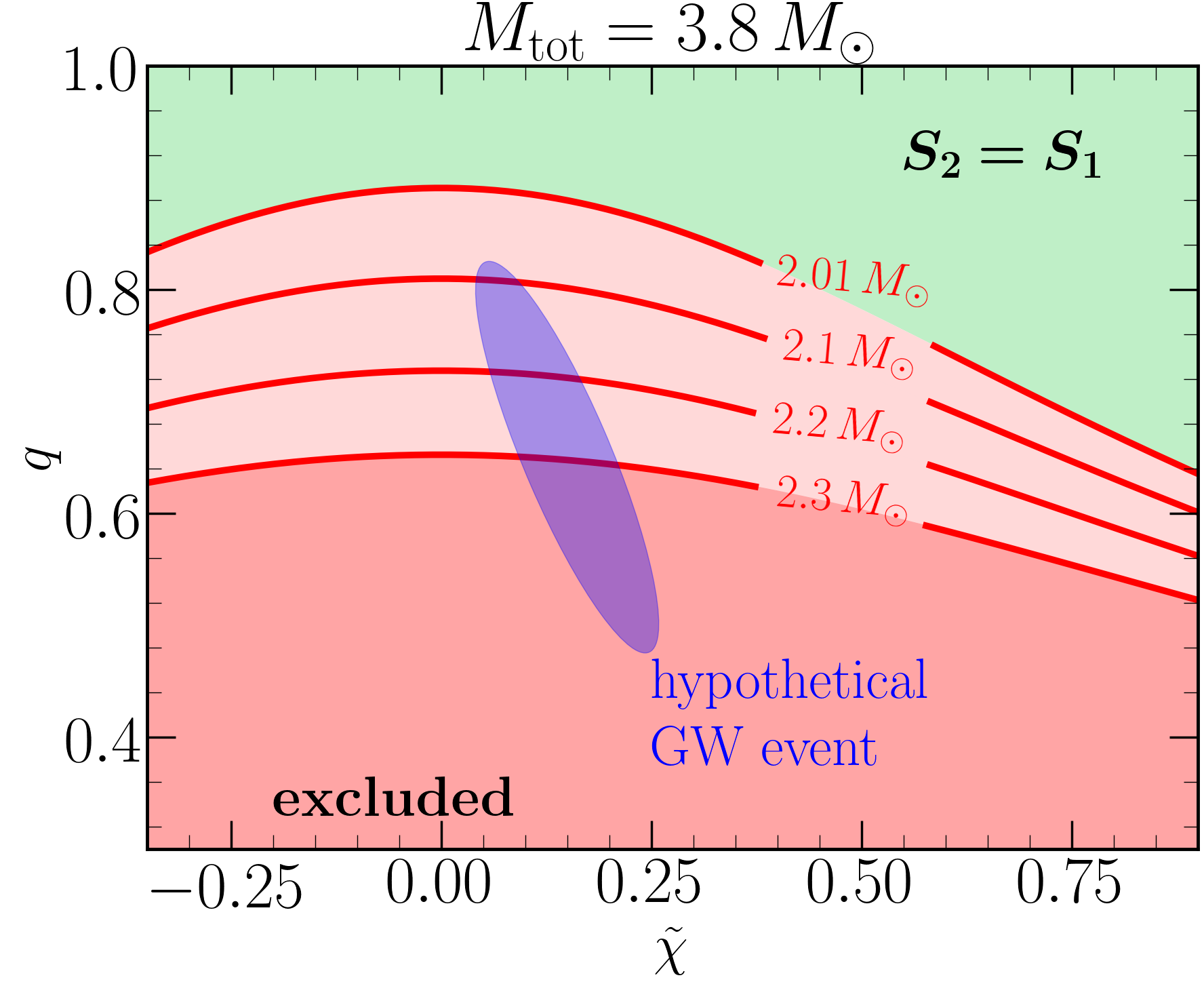}
  \caption{Same as Fig. \ref{fig:GW190425_const} but for a hypothetical
    future gravitational event. A large total mass and a large asymmetry
    in the mass ratio would a provide stringent lower bound on $M_{_{\rm
        TOV}}$.}
  \label{fig:future}
\end{figure}

\section{Conclusions and Outlook}
\label{sec:conclusion}

We have investigated how the properties of high-mass neutron-star
mergers, such as the mass ratio $q$ and the dimensionless spin
$\tilde{\chi}$, can be constrained by combining our current knowledge of
the EOS of nuclear matter and universal relations. In particular,
combining universal relations for the critical mass of uniformly rotating
neutron stars $M_\mathrm{crit}$ \citep{Breu2016} with a large dataset of
physically plausible EOSs constrained by the GW170817 event
\citep{Most2018,Weih2019}, we have been able to derive a lower limit for
the mass ratio $q_{\rm{min}}$ and an upper limit on the dimensionless
spin $\tilde{\chi}_{\rm max}$. While these limits depend on the spin
configuration of the binary and on the assumed maximum mass for
nonrotating configurations $M_{_{\rm TOV}}$, we could set constraints on
the properties of the gravitational-wave event GW190525 after considering
four representative cases of aligned and antialigned binaries and an
upper limit of $M_{_{\rm TOV}}=2.3\,M_{\odot}$. In this way, we have
concluded that $q_{\rm min} \geq 0.48$ and $\tilde{\chi}_{\rm max}\leq
0.16$ for GW190425, thus ruling out some of the most extreme binary
configurations. Future refinements on the value of $M_{_{\mathrm{TOV}}}$
will further constrain these limits. 

As an interesting outlook, we finally illustrate the hypothetical
scenario of a neutron-star merger in which the observations -- either
from the gravitational-wave detection or from the electromagnetic
counterpart (\ie amount of mass ejection, estimated life time, etc.) --
set an upper limit on the mass ratio, $q_{\rm max}$. This is illustrated
in Fig. \ref{fig:future}, which is the same as the top right panel of
Fig. \ref{fig:GW190425_const}, and where we have considered a
hypothetical high-mass merger with total mass $M_{\rm
  tot}=3.8\,M_{\odot}$, which has set a constraint in the
$(\tilde{\chi},q)$ plane (blue-shaded area). Should this be the case, it
would then be trivial to place a lower bound on $M_{_{\mathrm{TOV}}}$
from the intersection of the detection band with the contour lines shown
in Fig. \ref{fig:future}. More formally, given an upper value for the
mass ratio $q_{\rm max}$ -- and almost independently of the estimate for
$\tilde{\chi}$ -- it would be possible to invert numerically
Eq. \eqref{eqn:M_crit} and find therefore a strict lower limit for
$M_{_{\mathrm{TOV}}}$. In the limit of $\tilde{\chi} \rightarrow 0$, this
will have the trivial solution $M_{_{\mathrm{TOV}}} > M_\mathrm{tot} /
(1+q)$.

\section*{Acknowledgements}

It is a pleasure to thank Yizhong Fan and L. Jens Papenfort for useful discussions and
comments. ERM and LRW acknowledge support through HGS-HIRe. Support comes
in part from HGS-HIRe for FAIR; the LOEWE-Program in HIC for FAIR;
``PHAROS'', COST Action CA16214 European Union's Horizon 2020 Research
and Innovation Programme (Grant 671698) (call FETHPC-1-2014, project
ExaHyPE); the ERC Synergy Grant ``BlackHoleCam: Imaging the Event Horizon
of Black Holes'' (Grant No. 610058);

\bibliographystyle{mnras}

\bibliography{aeireferences}

\end{document}